\documentclass[12pt]{iopart} 
\usepackage{iopams}
\usepackage{psfig}
\usepackage{graphpap}

\newcommand{\chisq}{$\chi^{2}$}

\begin{document}  
\title{The Analysis of Cosmic Ray Data}
\author{K J Orford}
\address{Physics Department\\ Durham University\\Durham DH1 3LE, UK}
\maketitle 
%           MAIN SECTION                   INTRODUCTION                      MAIN SECTION
%           MAIN SECTION                   INTRODUCTION                      MAIN SECTION
%           MAIN SECTION                   INTRODUCTION                      MAIN SECTION
\section{Introduction}

A statement of statistical belief not uncommon in cosmic ray work is: "you need five sigmas to convince me". 
This has some justification, in that the history of cosmic rays contains many instances when a source 
or effect is claimed but not subsequently substantiated. 
Frequently this has been due to incorrect application of some statistical technique, often a failure to account 
fully  for the 'degrees of freedom'.

Most of the present body of statistical knowledge has been developed for specific problems, few of which occur in cosmic 
rays, although one of the most useful of texts\cite{kn:eadie} was produced for experimental particle physicists. 
The analyser of cosmic ray data has particular problems: cosmic ray data requires great effort in collection and they
are unlikely, once analysed, to be repeated. 
The numbers are frequently small and there are usually data missing and frequently there is significant contamination 
by noise. 
Ideally, a statistical measure should be developed specifically for each application. This is the only way in which 
all of the parameters of the experiment can be allowed for in the analysis. 
It is more usual for a general statistical tool to be applied, for example \chisq, which may not be optimal for the 
purpose, and for some experimental variables to be ignored.

The focus of this review will be on methods of determining the presence of a signal rather than estimating some parameters 
of the data.
The aim is to gather together the recent developments in methods of analysis of the temporal and spatial features of 
cosmic ray data, especially where the methods used are not 'traditional'. 

Several new methods have been published recently which depend on Bayesian ideas, and these ideas have been introduced 
before the description of the methods.

%           MAIN SECTION           ON/OFF COUNTS
%           MAIN SECTION           ON/OFF COUNTS
%           MAIN SECTION           ON/OFF COUNTS

\section{On/Off Counts}

\subsection{Introduction}

The subject of detecting the presence of a source in counting rate data, using off-source control data has appeared 
many times\cite{kn:eadie,kn:hearn,kn:omongain,kn:gibsona,kn:lima}. 
Despite these numerous airings, erroneous statistical significances are occasionally still being published.
In principle the question is easy to pose: if $N_{ON}$ counts are detected when an instrument is pointed at a source 
and there is also a background counting rate, and $N_{OFF}$ counts are detected when it is collecting background counts 
only under otherwise identical conditions, what is the likelihood that there is a genuine source?

A common treatment is to give for the significance of the excess counts:
\begin{equation}
N_{SIGMA}=\frac{N_{ON}-\alpha N_{OFF}}{\sqrt{N_{ON}+\alpha^{2}N_{OFF}}} \label{eq:dcxs}
\end{equation}
where $\alpha$ is the ratio of time on-source to time off-source, $t_{ON}=\alpha t_{OFF}$.
This is based on the supposition that the best estimate of the observed 'signal' is $N_{ON}-\alpha N_{OFF}$, 
the variances of the ON and OFF counts for a Poisson distribution are $N_{ON}$ and $N_{OFF}$ and the 
variance of the difference between $N_{ON}$ and $\alpha N_{OFF}$ is the weighted sum of the variances. 
The statistic used is $Student's$ $t$ which in the limit of large numbers is Gaussian. 
Since the distributions of $N_{ON}$ and $N_{OFF}$ are Poissonian, this expression should be used only if the numbers
of events is sufficiently large for a Gaussian approximation to Poissonian to be valid.

It is an example of only one type of statistic which could be used in $ON/OFF$ situations - a \emph{goodness-of-fit} 
statistic to determine whether the observed data could have arisen from an \emph{a priori} distribution.
Other statistics could have been used, for example \chisq, which in this instance would have one degree of freedom. 
Asymptotically they should have the same result, that is they both should reject or accept the null hypothesis
equally.
In these tests the \emph{null hypothesis} is that the observations were both samples from the same population and
that any difference arose merely by chance. 
There is no explicit alternative hypothesis, but an implicit one: that if the difference between the counts
was unlikely to be due to chance, it arose because of a genuine source, strength unspecified.

%          Subsection   LIKELIHOOD ANALYSIS
%          Subsection   LIKELIHOOD ANALYSIS
\subsection{Likelihood Analysis}

An optimal test exists for the intermediate case where there are two completely specified hypotheses: $H_{0}$: 
the \emph{null hypothesis} as described above, and $H_{1}$: a hypothesis involving another model, usually including 
a specific 'signal'. 
In this rare (in cosmic rays) case, the Neyman/Pearson theorem shows that the likelihood ratio is optimal for any 
distribution function for the errors.

In the more usual case, $H_{1}$ is not fully specified, but has one or more free parameters. 
The null hypothesis $H_{0}$ is that $N_{OFF}$ and $N_{ON}$ are both samples of the same population for which the source 
strength $S=0$.
The alternative hypothesis $H_{1}$ is that $N_{ON}$ contains an \emph{unknown} source component, $S>0$.
In this case there is no optimal test, except that for errors of the exponential family, such as a Gaussian, the 
likelihood ratio is expected to be near-optimal. 

The problem was discussed at length twenty five years ago by O'Mongain\cite{kn:omongain} and Hearn\cite{kn:hearn} but 
was not solved satisfactorily, at least in this field, until the maximum likelihood treatment of Gibson et al.
\cite{kn:gibsona} and Dowthwaite et al.\cite{kn:dowthwaite} and later by a similar treatment by Li and Ma\cite{kn:lima}. 
In these treatments the observed $ON$ and $OFF$ counts are due to (i) an unknown background $B$ plus an unknown source 
$S$ and (ii) the same unknown background $B$ alone. The likelihood ratio is maximised with respect to the possible 
source counts:
\begin{eqnarray}
\lambda & = & \left(\frac{P\left(N_{ON},N_{OFF}\mid S=0\right)}{P\left(N_{ON},N_{OFF}\mid S=N_{ON}-\alpha N_{OFF} 
\right)}\right) \nonumber
\\
        & = & \left[ \frac{\alpha}{1+\alpha}\left(\frac{N_{ON}+N_{OFF}}{N_{ON}}\right)\right]^{N_{ON}} \left[\frac{1}{1+\alpha}
\left(\frac{N_{ON}+N_{OFF}}{N_{OFF}}\right)\right]^{N_{OFF}}        \label{eq:lambda}
\end{eqnarray}
A standard result\cite{kn:eadie} is that the probability of obtaining a given $\lambda$ is obtained from
\begin{equation}
-2\ln(\lambda) \sim \chi^{2}(1) \label{eq:lamchi}
\end{equation}

%%%%%%%%%%%%%%%%%%%%%%%%%%%%%%%%%%%%%%%%%%%%%%%%%%%%%%%%%%%%%%%%
\begin{figure}[hb]
\psfig{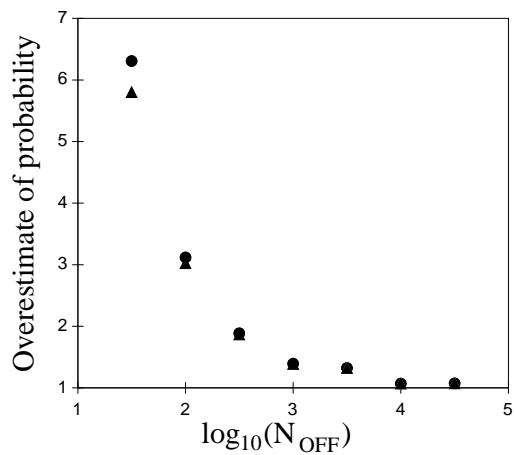}
\label{fig:1a}
\footnotesize
\caption{Ratio of probabilities from equation~\ref{eq:dcxs} (full circles) and equation~\ref{eq:lambda} (full triangles)
to Monte Carlo results for various values of $N_{OFF}$ and $N_{ON}=N_{OFF}+S_{3}$}     \label{fig:limacomp}
\end{figure}
\normalsize 
%%%%%%%%%%%%%%%%%%%%%%%%%%%%%%%%%%%%%%%%%%%%%%%%%%%%%%%%%%%%%%%%

%          Subsection   COMPARISON
%          Subsection   COMPARISON
\subsection{Comparison of methods}
Both equations~\ref{eq:dcxs} and~\ref{eq:lamchi} are valid asymptotically: only for large values of $N_{OFF}$ and $N_{ON}$.
Equation~\ref{eq:dcxs} assumes that the error distributions of $N_{ON}$ and $N_{OFF}$ are gaussian and equation~
\ref{eq:lamchi} assumes that $-2\log(\lambda)$ is distributed as $\chi_{2}(1)$.
To check the region of validity, random data sets have been generated for each of a number of values of $N_{OFF}$.
For each data set $\alpha$ has been set to $1.0$ and a value of $S_{3}=3\sqrt{2B+S_{3}}$ has been calculated, 
which is the $3\sigma$ value of $S=S_{3}$ assuming the validity of equation~\ref{eq:dcxs}.
At each value of $N_{OFF}$, $10^{6}$ data sets were generated using a poissonian random number algorithm\cite{kn:numreps}.
The fraction of samples $N$ of $N_{OFF}$ where $N > N_{OFF}+S_{3}$ is used as an estimate of the true probability of 
obtaining $N_{ON}=N_{OFF}+S_{3}$ by chance.
The results are shown in figure~\ref{fig:limacomp} where both equations~\ref{eq:dcxs} and \ref{eq:lamchi} are shown
to overestimate the probability near the $3\sigma$ level almost equally likely, the former slightly less so.
It is evident that, near the $3\sigma$ level, there is little to choose and both equations are adequate for values of 
$N_{OFF}$ and $N_{ON}$ of a few hundred or more. 
Since good algorithms are available for Poissonian random number generation it is likely to be better to determine
the probability of $N_{ON}$ and $N_{OFF}$ for values less than $\sim 100$ using Monte Carlo methods tailored for the 
exact values of $N_{ON}$ and $N_{OFF}$.

%           MAIN SECTION                   TIME SERIES                       MAIN SECTION 
%           MAIN SECTION                   TIME SERIES                       MAIN SECTION 
%           MAIN SECTION                   TIME SERIES                       MAIN SECTION 

\section{Time Series}

\subsection{Introduction}

Time series analysis has been the subject of very many books and articles and has been applied in very many fields. 
The term covers a wide range of concepts, including Change Point Analysis, Fourier Analysis and Trend Analysis.
In cosmic ray studies, there are several areas of application, such as sidereal/solar effects on low energy cosmic 
rays on the ground, periodicity in data from point sources, either from satellite X- and $\gamma$-ray data, 
or from ground-based \v{C}erenkov detectors, and sporadic emission of a wide range of cosmic ray energies. 
In these cases, the raw data is usually in the form of time-tagged events.

%           Sub Section           BURSTS
%           Sub Section           BURSTS

\subsection{Bursts of Events}

This section will be concerned with the problem of deciding whether the counting rate of a detector has deviated from 
the expected rate due to a real outburst of events.
The problem is usually most difficult in data comprising time-tagged events.
An initial analysis could start with binning the data and looking for a deviation from the expected Poissonian
distribution of the counts.
One problem with this approach is that in the model of a single Poisson process generating the counts, each bin is
independent, the experimenter often has the freedom to place the bins, both in position and width, arbitrarily.
This alters the 'degrees of freedom' and experience suggests that more bursts have been 'detected' in the past than
could have been justified from the data.

The problem mentioned above is a specific one but in general most statistical problems associated with sporadic emission 
relate to the lack of a specific model for the form, duration and amplitude of emission, and the feeling is often that, 
given a free hand with the parameters, any pure noise series could made to disclose a 'burst'.
A recent paper by Scargle\cite{kn:scargle98} suggests that existing methods for searching for rapid variability in $X-$ray 
and $\gamma$-ray astronomy do not fully extract all of the information contained in photon counts. 
The reasons given included 'binning fallacies', in that the data were widely binned and the size of the bins must be 
large enough to give 'good statistics'. 
Further, global methods such as autocorrelations and power spectra used on large data sets dilute the effects of sporadic 
bursts. The Bayesian response to these problems is discussed later. 

The problem at first sight does not seem insolvable using classical statistical theory. 
The statistical treatment of \emph{point processes}: data occurring as points on the real line, or as discrete times, 
is covered by several texts, for example Cox and Isham\cite{kn:cox}. 
The general treatment covers a variety of statistical processes, including Poisson (which is of most application here), 
doubly stochastic Poisson (where the average Poisson rate is itself a variable) and renewal processes where the distribution 
function for intervals between points is not exponential. 
In analysing data in the form of time-tagged photons without appreciable dead time, classical statistics would 
look for a powerful goodness-of-fit test of the pure Poisson process, if possible avoiding the loss of information and 
the arbitrary choices associated with binning.

Given such a series of times, the problem posed here is: is there evidence for 'bunching' or 'bursts'?
Alternatively, are the data consistent with a uniform distribution in time which, for events not affected by counter 
dead-time, would be governed by a pure Poisson process? 
Some recent papers such as McLaughlin et al.\cite{kn:mclaughlin} use just this assumption to classify sources into 'steady' 
or 'variable'. Others use \emph{ad hoc} methods to estimate the probability of bursts\cite{kn:katayose}.

\subsubsection{The Scan Statistic}

The test statistic postulated above exists: the Scan Statistic has been extensively studied by Parzen\cite{kn:parzen}, 
Barton and David\cite{kn:barton}, Huntington and Naus\cite{kn:huntington}, Neff and Naus\cite{kn:neff}, Naus\cite{kn:naus66},
Glaz\cite{kn:glaz93}, Wallenstein, Naus and Glaz\cite{kn:wallenstein73,kn:wallenstein94}, Chen and Glaz\cite{kn:chen97} and M\aa nsson
\cite{kn:mansson}.
It is a statistic for detecting clustering in time or one dimension in space. 
It is usually described as the maximum (or minimum) number of events which can be found in a window of fixed duration 
scanning smoothly through a much longer interval containing discrete events following some random process, for example 
Poissonian. 
An example of the scan statistic is shown in figure~1 for window lengths of 1\% and 10\% of the duration of the data. 
The random test data has a constant mean rate except for the third quarter which has double the rate.

%%%%%%%%%%%%%%%%%%%%%%%%%%%%%%%%%%%%%%%%%%%%%%%%%%%%%%%
\begin{figure}[h]
\psfig{file=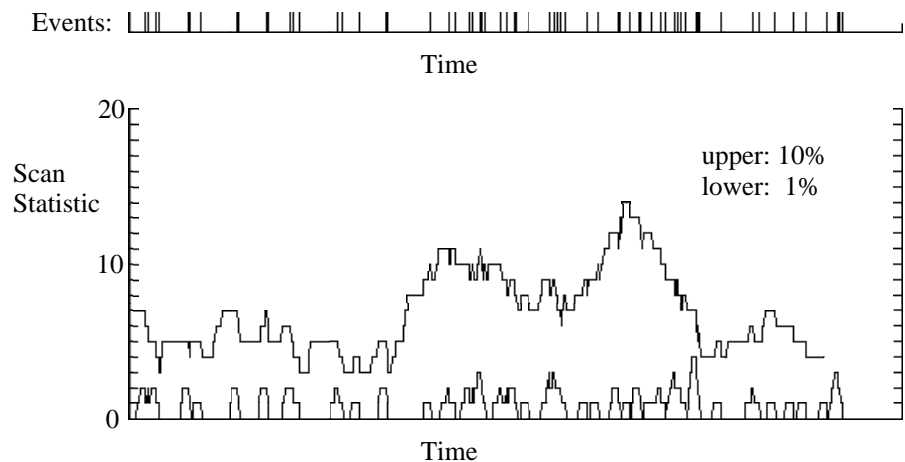,angle=90,height=6cm}
\label{fig:1b}
\footnotesize
\caption{Example of the Scan Statistic for windows of 1\% and 10\% of the data duration}
\end{figure}
\normalsize 
%%%%%%%%%%%%%%%%%%%%%%%%%%%%%%%%%%%%%%%%%%%%%%%%%%%%%%%

The scan statistic $S$ has a probability $P(S)$ which depends on the rate of events, the duration and the width of
the scanning window.
Some exact solutions for the probability $P(S)$ have been provided.
One of them, by Huntington and Naus\cite{kn:huntington}, provides the probability of a related statistic: 
\begin{equation}
P\left(a_{n}\leq a\right)=1-\sum_{Q}R \det\mid\!\!1/h_{ij}!\!\mid \det\mid\!\!1/l_{ij}!\!\mid
\label{eq:scanexact}
\end{equation}

where $a_{n}$ is the smallest interval $a$ containing $n$ events in the range $[0,1]$, where this range contains $N$ 
events in all. 
The summation extends over the set $Q$ of all partitions of $N$ into $2L+1$ integers satisfying $n_{i}+n_{i+1}<n, i=1,\ldots,2L$ and $R=N!b^{M}(a-b)^{N-M}$ with $M=\sum_{k=0}^{L}n_{2k+1}$, and 
\begin{eqnarray*}
h_{ij} & = & \;\;\;\sum_{k=2j-1}^{2i-1}\!\!n_{k}-(i-j)n\;\;\;\; L+1\geq i\geq j\geq 1
\\
       & = & -\sum_{k=2i}^{2j-2}\!\!n_{k}+(j-i)n\;\;\;\; 1\leq i<j\leq L+1
\\
l_{ij} & = & \;\;\;\sum_{k=2j}^{2i}n_{k}-(i-j)n\;\;\;\;L\geq i\geq j\geq 1
\\
       & = & -\sum_{k=2i+1}^{2j-1}\!\!n_{k}+(j-i)n\;\;\;\;1\leq i<j\leq L
\end{eqnarray*}
Equation~\ref{eq:scanexact}, although exact, is computationally expensive for large $N$ and small $a$, that is a large 
data set with a small scanning window, but several approximations have been provided which are designed to be valid for 
certain combinations of parameters.
 
\subsubsection{Newell-Ikeda Approximation for the Scan Statistic}

The Newell-Ikeda\cite{kn:newell,kn:ikeda} asymptotic formula is suitable for small probabilities. 
It gives the probability of finding a section of length $t$ in a data set of length $T$, given a Poisson process of 
average rate $\lambda$:
\begin{equation}
P\left( n;\lambda T,t/T \right) \doteq 1-\exp \left( -\lambda^{n}t^{n-1}T/(n-1)! \right)
\label{eq:newell}
\end{equation}
As shown in table~\ref{tab:scanstats}, it significantly overestimates larger probabilities.
\begin{table}
\begin{center}
\begin{tabular}{|l|l|l|} \hline
%\multicolumn{3}{l}{  } \\
$n$ & exact & Newell-Ikeda \\ \hline
$5$ & $0.711$ & $0.984$ \\ \hline             

$6$ & $0.225$ & $0.565$ \\ \hline
$7$ & $0.0425$ & $0.129$ \\ \hline
$8$ & $6.31\times10^{-3}$ & $0.0196$ \\ \hline
$9$ & $8.04\times10^{-4}$ & $2.48\times10^{-3}$ \\ \hline
$10$ & $9.05\times10^{-5}$ & $2.76\times10^{-4}$ \\ \hline
\end{tabular}
\caption{Comparison of Scan Statistic probabilities} \label{tab:scanstats}
\end{center}
\end{table}
Better approximations, although not as easy to calculate, are available, for example Conover, Bement and Iman
\cite{kn:conover} and Naus\cite{kn:naus82}. 

\subsubsection{Naus Approximation for the Scan Statistic}
The more exact treatment of Naus\cite{kn:naus82} will be given without derivation.
For an average rate of events $\lambda$, data of total duration $T$ and scanning window of duration $t$, define $L=T/t$. 
Then the probability that the number of events in a scanning window never exceeds $n$ is $Q^{*}\left(n;\lambda L,1/L\right)$
 and is accurately approximated by:
\begin{equation}
Q^{*}\left(n;\lambda L,1/L\right) \doteq Q^{*}\left(n;2\lambda,\frac{1}{2}\right)\left[Q^{*}\left(n;3\lambda,\frac{1}{3}
\right)/Q^{*}\left(n;2\lambda,\frac{1}{2}\right)\right]^{L-2}
\label{eq:naus}
\end{equation}
Note that this approximation is valid for a wide range of types of distribution for the time between events.
Exact formulae for $Q^{*}\left(n;2\lambda,\frac{1}{2}\right)$ and $Q^{*}\left(n;3\lambda,\frac{1}{3}\right)$ are given for 
a Poisson process:
\begin{eqnarray*}
Q^{*}\left(n;2\lambda,\frac{1}{2}\right) & = & F_{n-1}^{2}-\left(n-1\right)p_{n}p_{n-2}-\left(n-1-\lambda\right)p_{n}F_{n-3}
\\
Q^{*}\left(n;3\lambda,\frac{1}{3}\right) & = & F_{n-3}^{2}-A_{1}+A_{2}+A_{3}-A_{4}
\end{eqnarray*}
where
\begin{eqnarray*}
A_{1} & = & 2p_{n}F_{n-1}\left(\left(n-1\right)F_{n-2}-\lambda F_{n-3}\right)
\\
A_{2} & = & 0.5p_{n}^{2}\left(\left(n-1\right)\left(n-2\right)F_{n-3}-2\left(n-2\right)\lambda F_{n-4}+\lambda^{2}F_{n-5}
\right)
\\
A_{3} & = & \sum_{r=1}^{n-1}p_{2n-r}F_{r-1}^{2}
\\
A_{4} & = & \sum_{r=2}^{n-1}p_{2n-r}p_{r}\left(\left(r-1\right)F_{r-2}-\lambda F_{r-3}\right)
\end{eqnarray*}

and $p_{i}$,$F_{n}$ are the Poisson probability and distribution functions:  $p_{i}=e^{-\lambda t}(\lambda t)^{i}/i!$ and 
$F_{n}=\sum_{i=0}^{n}p_{i}$. 

Tight bounds for $Q^{*}(n)$ have been given by Glaz and Naus\cite{kn:glaz91} and a recursive method proposed\cite{kn:karwe} 
for calculating $Q^{*}(n;2t)$ and $Q^{*}(n;3t)$ for situations where the random quantity $X_{i}$ may take on values other 
than $0,1$, that is situations where an 'event' cannot be given as either present or absent but only with a non-zero 
probability.

Other approximations for the tail of the scan statistics and the moments of its distribution have been given by Glaz
\cite{kn:glaz93} and Chen and Glaz\cite{kn:chen97}. 
Sample tables of the scan statistic have been given for $n\leq 500$ by Glaz\cite{kn:glaz92,kn:glaz93}. 

This treatment of the scan statistic is for an interval of length $t$, specified in advance. 
When searching for a 'burst' of events, an \emph{a priori} length cannot always be specified. 
An extension to the treatment above has been described by Nargawalla\cite{kn:nargarwalla} in which the length need 
not be pre-assigned.

\subsubsection{Alm Approximation for the Scan Statistic}
A new approximation has been given recently by Alm\cite{kn:alm} which is accurate and easy to calculate for large values 
of $T/t$ and $\lambda t$. 
This treatment examines the distribution of upcrossings, that is occurrences where the number of
events in the scanning window increases by 1 as the window is moved. 
By separating these events into \emph{primary} and \emph{secondary} upcrossings, the dependence of the second type from the 
first (almost) independent events allows significant simplifications.
If each window of length $t$ were independent, the expected number of events would be $\lambda t$
with a Poisson probability function $F_{\lambda t}(n)$ and distribution function 
$p_{\lambda t}(n)$. 
The approximation based on the ideas above gives the simple modification:

\begin{equation}
P\left(N\geq n\right) = 1 - F_{\lambda t}(n)\exp\left[-\left(1-\frac{\lambda t}{n+1}\right)\lambda\left(T-t\right)p_{\lambda 
t}(n)\right]         
\label{eq:alm}
\end{equation}

Equation~\ref{eq:alm} has been tested for $\lambda t=40$ and $T/t=3600$ using 10000 Monte Carlo simulations.
The results are shown in table~\ref{tab:montecarlo} for $13 \leq n \leq 23$.
It can be seen that equation~\ref{eq:alm} is a good approximation within the sampling errors. 

\begin{table}
\begin{center}
\begin{tabular}{|l|l|l|} \hline
%\multicolumn{3}{l}{  } \\
$n$ & Monte Carlo & Equation~\ref{eq:alm} \\ \hline
$13$ & $0.9984$ & $0.9993$ \\ \hline
$15$ & $0.9429$ & $0.9478$ \\ \hline
$16$ & $0.6654$ & $0.6635$ \\ \hline
$17$ & $0.3094$ & $0.3087$ \\ \hline
$18$ & $0.1055$ & $0.1098$ \\ \hline
$19$ & $0.0307$ & $0.0337$ \\ \hline
$20$ & $0.0093$ & $0.0095$ \\ \hline
$21$ & $0.0021$ & $0.0025$ \\ \hline
$22$ & $0.0006$ & $0.0006$ \\ \hline
$23$ & $0.0002$ & $0.0001$ \\ \hline
\end{tabular}
\caption{Comparison of 1-D Scan Statistic probabilities for 10000 samples with $T/t=3600$ and $\lambda t = 40$} \label{tab:montecarlo}
\end{center}
\end{table}

%%%%%%%%%%%%%%%%%%%%%%%%%%%%%%%%%%%%%%%%%%%%%%%%%
%\begin{figure}[hb]
%\psfig{file=Almcomb.eps,height=6cm}
%\label{fig:11}
%\footnotesize
%\caption{Comparison of ratio: actual/approximate probability for various approximate formulae}
%\end{figure}
%\normalsize 
%%%%%%%%%%%%%%%%%%%%%%%%%%%%%%%%%%%%%%%%%%%%%%%%%

\subsubsection{Other Approximations for the Scan Statistic}
Other methods have been published, for example the Burst Expectation Search by Giles\cite{kn:giles} and CUSUM by 
VanStekelenborg and Petrakis\cite{kn:vanstekelenborg}. 
The first follows earlier work\cite{kn:rothschild} which used binned times of events and calculates Poisson probabilities 
of bin counts from a running average of a sample of bins. 
The BES inverts this process and, for each possible bin count from zero to several hundreds, calculates the mean rate 
below which the possible count could be a significant burst using a fixed sample of bins around the trial bin. 
The aim of keeping a fixed sample was to avoid problems arising from a step function edge entering a moving average.

\subsubsection{Bursts: Summary}
In summary, of possible methods suggested for searching for bursts using classical statistics, the Scan Statistic is 
recommended, both for time-tagged data and for time-binned data.
For small data sets or large window sizes, equation~\ref{eq:scanexact} provides an exact probability of the largest
number in any window arising due to chance.
Many approximate formulae are available, depending on whether the probability of the scan statistic is expected to 
be large or small.
In most practical cases in cosmic rays the statistic is used to search for a possible outburst and so the probability
of a given value of the scan statistic arising due to chance will be small in order to be useful.
It becomes a matter of computational convenience which of the formulae above is used but equation~\ref{eq:alm} delivers 
a good approximation over a wide range of probability values and is easy to calculate.
It also has the advantage, in terms of understanding the principles, of starting from the naive initial Poissonian 
formula with non-overlapping (independent) windows.
Its use is therefore recommended here.

%           Sub Section           PERIODICITY
%           Sub Section           PERIODICITY
\subsection{Periodicity}

Most of the methods for time-series analysis, including trend analysis and auto-regressive moving average (ARMA), 
have been developed for fields other than cosmic rays, for example \cite{kn:bell,kn:lathi,kn:lewis93}. 
Fourier methods suitable for data at equally spaced times are well developed but are usually not suitable, although 
these have been extended to discuss unequal intervals and missing data\cite{kn:scargle82}. 
A bibliography of astronomical time series analysis has been given by Koen\cite{kn:koen}.

\subsubsection{The Rayleigh test and Dependants}

The spur for the introduction of the Rayleigh test into $\gamma$-ray astronomy was the unsatisfactory nature of the 
statistics being used before. 
Early tests on $\gamma$-ray data used epoch-folding to produce a histogram in phase, and 
$\chi^{2}$ as a statistic for goodness of fit to a uniform distribution. 
This suffers from several disadvantages:
\begin{enumerate}
\item the freedom to select the number of bins, 
\item the freedom to define the starting phase, 
\item the failure to use the information contained in the order of the bins.
\end{enumerate}
This last problem can be overcome to some degree by using the Run Test, which is independent and therefore whose 
probability may be combined with that from \chisq. 
The result of the freedoms listed above is that different authors could return quite different chance probabilities, 
given the same data, despite using the same test statistic.
The analysis of $\gamma$-ray data from the Crab pulsar by Gibson et al.\cite{kn:gibsonb} contained the first known use 
of the Rayleigh statistic\cite{kn:mardia,kn:fisher} in cosmic ray work. 
It is still a goodness-of-fit statistic, which has no explicit hypothesis as an alternative to the null hypothesis.

The time of each event is treated as a unit vector in the plane, with an angle equal to the pulsar phase. 
If $N$ unit vectors of random orientations (random phases) are added, the distribution  of the resultant $R$ may  
be obtained from the distribution of the orthogonal components of the vectors, $\sin \phi_{i} $ , $\cos \phi_{i}$
where $\phi_{i}$ is the phase of the $i^{th}$ vector. 
The means of these components are :
\begin{displaymath}
C = \frac{1}{N}\sum_{i=1}^{N} \cos \phi_{i}  \ \ \ \ \ S = \frac{1}{N}\sum_{i=1}^{N} \sin 
\phi_{i}
\end{displaymath}

From the Central Limit Theorem (CLT) means of samples of $C$ are distributed, for large $N$, as a Gaussian with $var_{C} = 
\sigma_{C}^{2}/N$.
For vectors uniformly covering the circle:
\begin{displaymath}
  \sigma_{C}^{2} = \int_{0}^{2\pi} \cos^{2}(\phi) d\phi / 2\pi = 0.5
\end{displaymath}
 therefore $var_{C} = var_{S} = 1/2N$.
The quantities $C$ and $S$ are asymptotically uncorrelated and have zero means. 
The statistic \(2NR^{2} = 2NC^{2}+2NS^{2}\) is therefore the sum of the squares of two zero-mean, unit-variance 
uncorrelated variables and is distributed as $\chi^{2}$ with 2 degrees of freedom \cite{kn:priestley, kn:bloomfield}. 
The probability distribution function (pdf) of $R$ is :  
\begin{equation}
%*}
  f(R)dR = 2NR e^{-NR^{2}} dR   
\label{eq:pdf}
\end{equation}
%*}
and its cumulative probability distribution is :
\begin{equation} 
  F(R) = e^{-NR^{2}}    \label{eq:probab}
\end{equation}
The quantity $NR^{2}$ is known as the Rayleigh power.

If a data set spans a time interval $T$ the number of independent frequency trials in the frequency range $f_{1}$ to $f_{2}$
 is $\nu=T\left(f_{1}-f_{2}\right)$ if $T >> f_{1}$,$f_{2}$, with the independent frequencies separated by $1/T$. 
In practice allowance must be made for leakage: the possible effect of a signal at frequency $f_{0}$ on trial frequencies 
$f$ with $\mid \!f-f_{0}\mid\!>1/T$, and oversampling: the possibility of obtaining a larger value of $NR^{2}$ by
varying the frequency between adjacent independent frequencies. 
This has been done using by de Jager et al.\cite{kn:dejager89} by Monte Carlo techniques and analytically by 
Orford\cite{kn:orford91}. 
Both methods agree that the number of trials is $nT\left(f_{1}-f_{2}\right)$ where $n$ is a slowly 
varying function of $F(R)$ in the range $2$ to $4$, with a value approximately $3$ for $F(R) \sim 10^{-3}$.

%        THE Zn TEST
\subsubsection{The $Z_{n}^{2}$ Test}

The $Z_{n}^{2}$ test is the extension of the Rayleigh test to include harmonics.
IF \textbf{n} separate harmonics are included with independent coefficients, the statistic is 
\begin{equation}
Z_{n}^{2}=2N\sum_{i=1}^{n} R^{2}(i\omega)   \label{eq:zn}
\end{equation}
where $2NR^{2}(i\omega)$ is the Rayleigh power for the $i^{th}$ harmonic.
$Z_{n}^{2}$ is distributed as $\chi^{2}(2n)$.
Variations on this technique depend on the method used to select the number and weighting of harmonics. 
A similar principle is used in radio astronomy where a pulse of width $W$ is searched for using $P/2W$ harmonics 
which improves the signal to noise by a factor of up to $(P/2W)^{0.5}$\cite{kn:lyne}.
A search for $\gamma$-ray emission from radio pulsars proposed the use of $Z_{2}^{2}$ as a relatively powerful but
general test for periodicity\cite{kn:buccheri}.
The power of the Rayleigh test for light curves from sinusoids to $\delta$-functions was explored by Protheroe
\cite{kn:protheroe3}.
A variant of $Z_{n}^{2}$ is the H-test\cite{kn:dejager89} in which the value of $n$ is obtained objectively from the data
and $Z_{n}^{2}$ is suitably rescaled.
This last test is most suitable for multi-mode light curves.

%        RED NOISE IN RAYLEIGH TESTING
\subsubsection{Limitations of the Rayleigh and associated statistics}

The foregoing results for the Rayleigh ($Z_{1}^{2}$) and $Z_{n>1}^{2}$ tests are for the asymptotic case, that is: uncorrelated 
$C$ and $S$ with zero means. 
In most practical applications, these conditions are not strictly met. 
Ground-based gamma-ray observations of long-period pulsars are limited by: 
\begin{enumerate}
\item being only a few hours in duration and 
\item variations in zenith angle, producing changing counting rates.
\end{enumerate} 
The requirement for large sample size is usually met - typical counting rates are $\approx$~1 per second over several hours. 
The result is an enhancement of \chisq\ in pure noise data for longer test periods - red noise.
The first limitation listed above may be overcome by truncating the dataset so that only integral multiples of the trial 
period are tested - see Carrami\~{n}ana et al.\cite{kn:alberto} and Raubenheimer and \"{O}gelman\cite{kn:raub1}. 
As a result, the two trigonometric terms have zero expectations, given a constant mean counting rate. 
This truncation is easy to accomplish, but results in a variable data selection depending on the test period and therefore 
all periods are not accorded the same treatment. 
Since the periodogram is the convolution of the power spectral density with the Fourier transform of the data window, any 
spectral estimate based on a truncated data set is biased\cite{kn:kay,kn:priestley,kn:jenkins}. 
Further, any correlation introduced by the second limitation above will not be removed this way. 
An attempt to remove the results of the counting rate variation has been made by Raubenheimer et al.\cite{kn:raub2} by 
fitting a parameter $a$ in an \emph{ad hoc} modification of the Rayleigh probability distribution: 
\begin{equation}
  F(R) = e^{-2aNR^{2}} 
\label{eq:raubenheimer}
\end{equation}
to random data sets containing no signal, but with the same parameters as the test data set. 
For data taken on Vela X-1 (period $\approx$~ 5 minutes) they found that equation~\ref{eq:raubenheimer} with $a=0.4$ 
(as opposed to 0.5 from simple theory) gave a probability distribution which was a good fit to the distribution in 
\chisq\ for noise at periods near to 5 minutes in simulated data sets.
%         MODIFIED RAYLEIGH STATISTIC

\subsubsection{Modified Rayleigh Statistic}

If the expectations of $C$ and $S$, their variances and their covariance are not assumed to be zero, $\frac{1}{2N}$ and 
zero respectively, but are calculated for a specific dataset, then the asymptotic probability equation~\ref{eq:probab} 
may be valid, given a sufficiently large number of events\cite{kn:orford96}. 

The expression for \chisq\ in the case of samples of two correlated variables $C$ and $S$ is :
\begin{eqnarray} 
  \chi^{2}&=&
  \left[
  \begin{array}{c} 
    \overline{C}-E(C) \\ \overline{S}-E(S) 
  \end{array} 
  \right] ^{T}
  \left[ 
  \begin{array}{cc} 
    \sigma_{C}^{2} & cov_{S,C} \\ cov_{S,C} & \sigma_{S}^{2} 
  \end{array} 
  \right] ^{-1} 
  \left[ 
  \begin{array}{c} 
    \overline{C}-E(C) \\ \overline{S}-E(S) 
  \end{array} 
  \right] \label{eq:chisq}
\end{eqnarray}

For any data set, the substitution of the actual values of $E(C)$, $E(S)$, $\sigma_{C}$, $\sigma_{S}$ and $cov_{S,C}$ will 
result in a value of \chisq\ corrected for the correlation of the variables $C$ and $S$ and with a probability distribution,
for large sample size, given by $\exp(-\chi^{2}/2)$.

In the case of a box-car data set with a constant average counting rate, a starting time $t_{1}$ and ending time $t_{2}$ 
with $T=t_{2}-t_{1}$ and a trial period   $P = 2\pi/\omega$ :
\begin{eqnarray*}
 E(C) & = & \frac{\left[\sin\omega t\right]_{t_{1}}^{t_{2}}}{\omega T} \label{eq:exc} 
\\
 E(S) & = & \frac{\left[-\cos\omega t\right]_{t_{1}}^{t_{2}}}{\omega T} \label{eq:exs}
 \\
 Nvar_{C} & = & \frac{1}{2} + \frac{\left[\sin \omega t \cos \omega t\right]_{t_{1}}^{t_{2}}}{2\omega T} - [E(C)]^{2}
\label{eq:varc}
 \\ 
 Nvar_{S} & = & \frac{1}{2} - \frac{\left[\sin \omega t \cos \omega t\right]_{t_{1}}^{t_{2}}}{2\omega T} - [E(S)]^{2}
\label{eq:vars}
 \\
 Ncov_{S,C} & = & \frac{\left[\sin^{2}\omega t\right]_{t_{1}}^{t_{2}}}{2\omega T}  -  {E(C)E(S)} \label{eq:covsc}
\end{eqnarray*} 
These depend solely on $\omega$, $t_{1}$ and $t_{2}$ and their substitution in equation~\ref{eq:chisq} gives a \chisq\ 
value corrected for the finite length of the data set. 
If it is known that there is no secular change in counting rate the substitution of the above equations 
into equation~\ref{eq:chisq} would give the correct formal probability of chance occurrence, even if the duration of the 
data set is less than the trial period, as long as the number of events was high enough for the CLT to be valid. 
It is more usually the case in ground-based gamma-ray observations that the box-car function is only an approximation.
Monte Carlo simulations of data sets have been carried out to test the validity of data set truncation and the above 
formulations for the case of secular variations of counting rate superimposed on noise. 
In order to test the validity of the probability distribution equation~\ref{eq:probab} down to probabilities 
$\approx 10^{-6}$, data sets were generated using a multiplicative congruential algorithm with shuffling, chosen to 
avoid serial correlations. 
The repeat period is longer than $2\times10^{18}$. 
The time of each event $t_{i}$ was generated from the previous event:
$t_{i}=t_{i-1}-\Delta(t)\ln(rnd)$ where $\Delta(t)$ is the mean separation of events as a function of time.
A group of $10^{6}$ data sets of duration 8000 s were simulated with a counting rate profile \mbox{$R=R_{0}(1-0.3t/4000)$} 
and \mbox{$R_{0}=1s^{-1}$}. 
Each data set was tested for periodicity at a trial period of 295s by finding $C$ and $S$ with reference to the time 
of the first event.
 These values were substituted into equation~\ref{eq:chisq} for various assumptions about the form of $E(C)$, $E(S)$, 
$var_{C}$, $var_{S}$ and $cov_{S,C}$. The probability of chance occurrence was calculated from $e^{-\chi^{2}/2}$.

%%%%%%%%%%%%%%%%%%%%%%%%%%%%%%%%%%%%%%%%%%%%%%%%%%%%
%\begin{figure}[h]
%\begin{center}
%\psfig{file=Fig2.eps,angle=90,height=6cm}
%\footnotesize
%\caption{Frequency distribution of \chisq\ probabilities from $10^{6}$ Monte Carlo data sets} \label{fig:monte2}
%\end{center}
%\end{figure}
%\normalsize 
%%%%%%%%%%%%%%%%%%%%%%%%%%%%%%%%%%%%%%%%%%%%%%%%%%%%

\begin{figure}[ht]
\center
\psfig{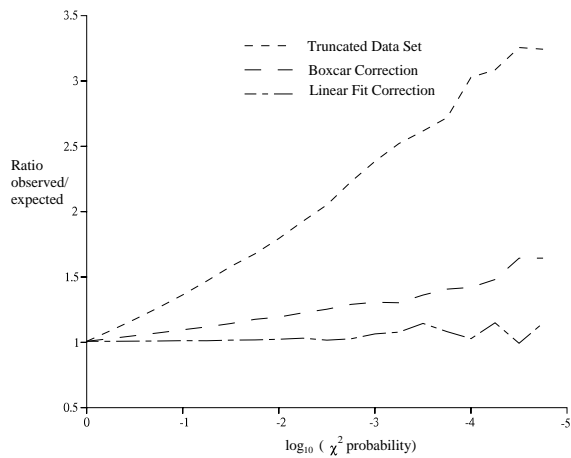}
\footnotesize
\caption{Ratio of cumulative frequency of \chisq\ probabilities to expectation} \label{fig:cumfreq}
\end{figure}

The resulting cumulative frequency distributions for $e^{-\chi^{2}/2}$
% are shown in figure~\ref{fig:monte2} 
have been calculated for the cases 
of (a) a truncation of the data set to integral multiple of the trial period, (b) box-car 
function and (c) a linear fit to the counting rate profile. 
The ratios of the observed to expected frequencies of occurrence of \chisq\ chance probabilities is shown in 
figure~\ref{fig:cumfreq}, as functions of $\log{(\chi^{2} probability)}$.
Note that the duration of the data set is corrected for equally well by (b) and (c).

%%%%%%%%%%%%%%%%%%%%%%%%%%%%%%%%%%%%%%%%%%%%%%%%%%%%
\normalsize
%%%%%%%%%%%%%%%%%%%%%%%%%%%%%%%%%%%%%%%%%%%%%%%%%%%%

The boxcar and truncated statistics both make corrections for the finite length of a data set, but give a residue which 
may be identified as being caused by the change of counting rate during the trial period. 
Longer trial periods or greater rates of change in counting rate would amplify their biases. 
The truncation method has a distribution which may be fitted, for this simulated data set, by a form such as 
equation~\ref{eq:raubenheimer} with $a=0.45$. 
The linear fit model is seen to be a good representation of the noise spectrum down to chance probabilities of 
$\approx 10^{-5}$.
%               OTHER TESTS
%               OTHER TESTS
\subsubsection{Other Tests}

Leahy et al.\cite{kn:leahy} pointed out that the unmodified Rayleigh test was powerful for detecting wide peaks in a 
light curve, in fact it is identical with a likelihood ratio test of a sinewave plus uniform against a uniform phase 
distribution \cite{kn:bai}. 
In addition, for a light curve of a von Mises form (the circular generalisation of the Gaussian), the Rayleigh statistic
exhausts the data's information on periodicity if the concentration parameter $\kappa$ is allowed to vary freely
\cite{kn:loredo92}.

Narrow periodic pulse detection, with significant power in the higher harmonics, is bound to be quite difficult because 
the number of degrees of freedom increases with the trial frequency range. 
Protheroe\cite{kn:protheroe1} proposed a test statistic 
\begin{equation}
T_{n} =\frac{2}{n(n-1)}\sum_{i=1}^{n-1}\sum_{j=i+1}^{n}(\Delta_{ij}+1/n)^{-1}
\end{equation}
which looked for close clustering of points on the circle. In this statistic $\Delta_{ij}$ is the distance between the 
angles $x_{i}$ and $x_{j}$ of two events on the circle:
\begin{displaymath}
\Delta_{ij}=0.5-\mid\left[\mid(x_{i}-x_{j})\mid-0.5\right]\mid
\end{displaymath}
The null distribution was found using Monte Carlo methods for $n\leq200$ and critical values given. 
The context of the test was the search for ultra-high energy $\gamma$-rays from Cygnus X-3 and was therefore not 
designed for large $n$. 
In this limitation it is similar to the exact expression scan statistic described above. 
Others have suggested variants which are designed to be powerful for certain classes of pulsed emission
\cite{kn:swanepoel}.

The Scan Statistic may also be used for searching for non-uniformity in phase. 
For narrow windows its probability distribution is well approximated by the scan statistic on the line\cite{kn:nauspriv}.
No systematic work on the use of the Scan Statistic in periodic analysis has been traced.
%        SEARCHING FOR PERIODICITY
%        SEARCHING FOR PERIODICITY
\subsubsection{Searching for a Periodicity}

It is usually only the case that a unique periodic ephemeris is available for high energy photons from an isolated radio 
pulsar.
In other cases, a search must be made in period, and the test used must allow for the freedoms associated with the
trial period range.
A rule-of-thumb arising from the number of 'degrees of freedom' implicit in a periodicity search using the Rayleigh
test\cite{kn:dejager89,kn:orford91} is that the search should be at intervals in period of $(IFI)/3$ ($IFI$ = 
Independent Fourier Interval).
For a period of $P$ in a data set of duration $T$ this corresponds to a trial period step of $\sim P^{2}/3T$.
This step size has the advantage that the number of degrees of freedom to be used to interpret the peak periodic amplitude
found is approximately the number of periods tried.
If harmonics of the test period are to be included, the spacing would be correspondingly reduced and hence the number of
trial periods increased.
For the $Z_{n}^{2}$ test, the reduction in period step, and consequent increase in both computation and the degrees of
freedom to be accounted for, is by a factor $n$.

When searching for pulsed emission from some sources, in particular binary sources, there is frequently poor knowledge of 
the both the pulsar period and period derivative. 
In this case the light curve will be narrow only if the correct period $P$ and period derivative $\dot{P}$ is offered to 
the test.
A nearby, but not correct, trial period and the ignorance of a period derivative will smear the light curve.
If a true light curve were a $\delta$ function at period $P=P_{o}$ and the trial period was $P=P_{o}+\Delta$, the light
curve would be a rectangular distribution in phase of length $T\Delta /P_{o}^{2}$.
This effectively limits the number of harmonics which may be realistically added to $P_{o}^{2}/T\Delta-1$.

Some searches for periodicity are combined with a search for a DC excess. This is common in \v{C}erenkov telescope 
searches where ON-source data is compared with OFF-source control data to detect any DC component. 
The combined analysis of this situation was proposed by Lewis\cite{kn:lewis89} in which a statistic $\alpha$ is defined as the 
sum of the Rayleigh statistic and the square of equation~\ref{eq:dcxs}, distributed as \chisq(3). 
The assumption in this case is that all of the excess is pulsed; if there is an unpulsed component the test statistic 
will be biased. 
Again, the presence of a possible unpulsed component could be built into a Bayesian analysis.
%    CONCLUSION
\subsubsection{Conclusion on Periodicity}
The question of the best classical test for the presence of periodicity is a complex one.
The selection of the most sensitive test requires a knowledge of 
\begin{itemize}
\item the pulsar ephemeris
\item the light curve shape
\item the background noise distribution
\end{itemize}
If all of these are known in advance, a most powerful test, based on the $Z_{n}^{2}$ extension of the Rayleigh test
is likely to be close to optimum.
Frequently some or all of these will be unknown or poorly known.
In this case, some allowance must be made for the lack of knowledge and the test selected should not contain any
assumption which causes a significant bias.
It has sometimes been claimed that the Rayleigh test is 'biased' towards broad light curves and that a test which
is more sensitive to narrow light curves should be used when such a light curve is suspected.
This raises the problem, discussed in the previous section, of the smearing of a light curve if the pulsar's ephemeris is
uncertain. 

Protheroe has suggested\cite{kn:protheroe2} that if one has no information about the nature of the phase distribution 
one should be conservative and adopt the Rayleigh test. 
A rational for this is that if one is searching for an unknown period and an unknown light curve, which is quite common 
in $\gamma$-ray work, and there is no significant power in the fundamental, then a test involving the addition of an 
unknown number of higher harmonics is unlikely to be successful.
This point will be revisited later in discussing a Bayesian method of searching for periodicity.

A simple suggestion made before and reiterated here is that if $Z_{n}^{2}$ does not show evidence for periodicity,
that is: there is no significant power in the fundamental or the first harmonic, either in addition or separately, then it 
is unlikely that the data will contain a strong periodic signal.

%           MAIN SECTION                   SPATIAL ANALYSIS                      MAIN SECTION 
%           MAIN SECTION                   SPATIAL ANALYSIS                      MAIN SECTION 
%           MAIN SECTION                   SPATIAL ANALYSIS                      MAIN SECTION 

\section{Spatial Analysis}

\subsection{Introduction}  

Spatial analysis of arrival direction data is of great interest for X- and $\gamma$-rays from satellites and for 
cosmic rays of the highest energies, which may not be greatly deflected in the galactic magnetic field. 

Simple methods rely on a grid placed on the events and counts in the grid cells taken as independent Poisson-distributed 
events.
If the cells are fixed absolutely, there is no problem in ascribing a suitable Poissonian probability to the largest
number detected in any cell.
If there is freedom to incrementally move the cell containing the largest count, a larger number is generally found.
In this case the new cells created are correlated and the assumption of independence is incorrect: simple application
of Poissonian probabilities is inappropriate.
The problem of having the freedom to move the boundaries of the cells was pointed out for cosmic ray 'sources' by 
Hillas\cite{kn:hillas} who suggested a conservative number of 'sigmas'.
Large scale anisotropy in gamma ray bursts were sought using dipole and quadrapole analysis\cite{kn:briggs}. 
A 'pair matching' statistic was used by Bennett and Rhie\cite{kn:bennett} to check for gamma ray burst repeaters rather 
than 'nearest neighbour' methods used by others\cite{kn:brainerd} and criticised by Nowak\cite{kn:nowak}.
Many methods have been used which are based upon a known point-spread function (PSF). 
Amongst these are Maximum Entropy methods such as those used for satellite X-ray imaging\cite{kn:cheng}, maximum
likelihood\cite{kn:mattox} and Hough Transforms\cite{kn:ballester}. 

In the next section it is suggested that the scan statistic is a powerful and general statistic for which good approximations
exist for the chance probabilities.
It has recently been extended to two dimensions by Loader\cite{kn:loader}, Chen and Glaz\cite{kn:chen96} and 
Alm\cite{kn:alm}. Kulldorf\cite{kn:kulldorff} has extended this further to higher dimensional searches.

%           sub section   2-D SCAN STATISTIC
%           sub section   2-D SCAN STATISTIC
\subsection{2-Dimensional Scan Statistic}
This is the two-dimensional development of the Scan Statistic introduced above.
It will be introduced using a notion of 'elemental' cells from which a two-dimensional scanning window is constructed.
In effect, the scanning window may be moved by discrete steps of the size of the elemental cell.
Assume that a two-dimensional square region $R=\left[0,L\right]\times\left[0,L\right]$ of side $L$ is inspected for 
occurrences of 'sources'\cite{kn:chen96}. 
The region is partitioned into $n\times n$ elemental cells so that the size of a cell $h=L/n$. 
The contents of each of the $n^{2}$ cells are independent.

%%%%%%%%%%%%%%%%%%%%%%%%%%%%%%%%%%%%%%%%%%%%%%%%%%
\begin{figure}[ht]
\psfig{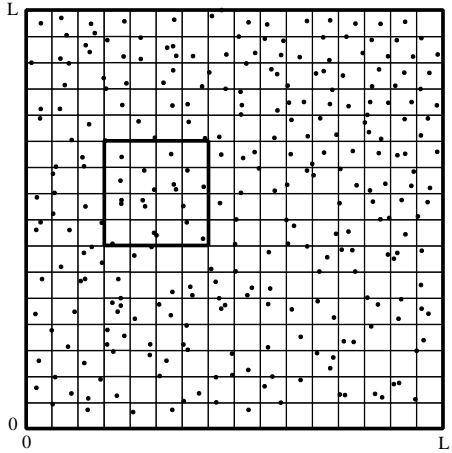}
\label{fig:2dscan}
\footnotesize
\caption{Two-dimensional Scan.
The region [0,L]X[0,L] is partitioned into $n\times n$ cells ($n=16$) and an $m\times m$ window ($m=4$) is scanned over it.}
\end{figure}
%\end{picture}
\normalsize
%%%%%%%%%%%%%%%%%%%%%%%%%%%%%%%%%%%%%%%%%%%%%%%%%% 

For $1\leq i\leq n$ and $1\leq j\leq n$, define a random variable $Y_{i,j}$ as the number of events in the elemental cell 
$\left[(i-1)h,ih\right] \times \left[ (j-1)h,jh\right] $. 
A square box of $m\times m$ small cells is scanned over the whole of region $R$.
There will be $\nu$ such boxes, partially dependent if $m \geq 2$, with $\nu = 0, (n-m+1)^{2}$.

Define \[ S(i_{1},i_{2})=\sum_{j=i_{2}}^{i_{2}+m-1}\sum_{i=i_{1}}^{i_{1}+m-1}Y_{i,j} \]
to be the number of events in the square box of $m^{2}$ adjacent cells starting at $i=i_{1}$, $j=i_{2}$. 
If, during the scanning of the $m-$box, $S(i_{1},i_{2})$ exceeds a particular value $k$, a 'source' has been detected.
For $1\leq i_{1},i_{2}\leq n-m+1$ define an 'event' $A_{i_{1},i_{2}}$ as an occurrence of $S(i_{1},i{_2})\geq k$ and 
as a member of the set $A$ of all such occurrences.

The two-dimensional scan statistic is defined as:
\[S_{m}=\max\left\{S(i_{1},i_{2});1\leq i_{1}\leq n-m+1,1\leq i_{2}\leq n-m+1\right\}\]
and the probability that $S_{m}$ has at least a value $k$ is:
\begin{eqnarray*}
P(S_{m}\geq k) & = & P\left(\bigcup_{i_{1}=1}^{n-m+1}\bigcup_{i_{2}=1}^{n-m+1}A_{i_{1},i_{2}}\right)
%\\             & = & P\left(A_{1,i_{2}} or A_{2,i_{2}} or ..... or A_{n+m-1,i_{2}}\right)
\\
               & = & P\left(\bigcap_{i_{1}=1}^{n-m+1}\bigcap_{i_{2}=1}^{n-m+1}A_{i_{1},i_{2}}^{c}\right)
\end{eqnarray*}
where $A_{i_{1},i_{2}}^{c}$ is the occurrence of $S(i_{1},i_{2}) < k$.

\subsubsection{Glaz Approximation to 2-D Scan Statistic}
For a fixed value of $1\leq i_{1}\leq n-m+1$ the one-dimensional approximation holds:
\[P\left(\bigcap_{i_{2}=1}^{n-m+1}A_{i_{1}i_{2}}^{c}\right)\approx q_{2m}\left(\frac{q_{2m}}{q_{2m-1}}\right)^{n-2m}\]
and since $n-m+1$ square regions of $m\times m$ are scanned, a reasonable approximation is\cite{kn:glaz91}:
\begin{equation}
P(S_{m}\geq k)\approx 1-q_{2m-1}\left(\frac{q_{2m}}{q_{2m-1}}\right)^{(n-2m+1)(n-m+1)}
\label{eq:2dapprox}
\end{equation}
For a Poissonian distribution of events the following expression was found to be a good approximation\cite{kn:chen96}:
\begin{equation}
P(S_{m}\geq k)\approx 1-\exp(-\lambda^{*})  \label{eq:2dapproxb}
\end{equation}
where the approximate mean for the asymptotic Poisson distribution is
\[\lambda^{*}=1-q_{2m-2}+(n-2m+2)(n-m+1)(q_{2m-2}-q_{2m-1})\]
and
\[q_{m+l-1}=P(A_{1,1}^{c}\cap A_{1,1}^{c}\ldots \cap A_{1,l}^{c})\]
Tables are given\cite{kn:chen96} of this and other approximations, for the Poisson model of $m\leq 20$ and $n\leq 500$.

\subsubsection{Alm Approximation to 2-D Scan Statistic}
A recent paper\cite{kn:alm} has given an approximation based on a modification of the method
of counting \emph{upcrossings} used in equation~\ref{eq:alm}, which is easy to calculate and 
moreover is given for a more generally useful rectangular scanning window $[0,a]\times [0,b]$ in a 
rectangular region $[0,S]\times [0,T]$. 
The scan statistic $L$ is the maximum content of a scanning window with a two-dimensional
Poissonian process $X$ with event density $\lambda$:
\[L=L(\lambda,a,b,S,T)=\max_{a\leq t\leq T, b\leq s \leq S}X([t-a,t]\times [s-b,s]\]
The probability of observing at least $n$ events in a scanning window is:
\begin{equation}
P(L\geq n) \approx 1 - F_{N(a)} e^{-\gamma_{n+1}}  \label{eq:alm2d}
\end{equation}
where
\[\gamma_{n+1} \approx \left(1-\frac{\lambda ab}{n+1}\right)\left(T-a\right)b\lambda \left(\mu_{n}-\mu_{n+1}\right)e^
{-\mu _{n+1}}\]

\[\mu_{n} \approx \left(1-\frac{\lambda ab}{n}\right)\lambda a\left(S-b\right)p_{\lambda ab}(n-1)\]
and
\[F_{N(a)} \approx F_{\lambda ab}\exp\left[1-\left(1-\frac{\lambda ab}{n+1}\right)\lambda a\left(S-b\right)p_{\lambda ab}(n)
\right]\]
$p_{\mu}$ and $F_{\mu}$ are the Poisson probability and cumulative probability distributions. 

\begin{table}
\begin{center}
\begin{tabular}{|l|l|l|l|} \hline
%\multicolumn{4}{l}{  } \\
$n$ & Monte Carlo & Equation~\ref{eq:alm2d} & Poisson \\ \hline
$ 7$ & $0.9990$ & $0.9975$ & $3\times 10^{-3}$\\ \hline
$ 8$ & $0.8658$ & $0.8795$ & $10^{-3}$ \\ \hline
$ 9$ & $0.4271$ & $0.4501$ & $2\times 10^{-4}$ \\ \hline
$10$ & $0.1351$ & $0.1346$ & $4\times 10^{-5}$ \\ \hline
$11$ & $0.0326$ & $0.0304$ & $7\times 10^{-6}$ \\ \hline
$12$ & $0.0089$ & $0.0059$ & $10^{-6}$ \\ \hline
$13$ & $0.0021$ & $0.0011$ & $2\times 10^{-7}$ \\ \hline
$14$ & $0.0005$ & $0.0002$ & $3\times 10^{-8}$ \\ \hline
\end{tabular}
\caption{Comparison of 2-D Scan Statistic probabilities for 100000 samples with N=800, S=T=40, a=b=2. Also shown are the
simple Poissonian probabilities assuming independent cells} \label{tab:2dscan}
\end{center}
\end{table}

The predictions of equation~\ref{eq:alm2d} have been compared with the results of $10^{5}$ Monte Carlo simulations in 
table~\ref{tab:2dscan} for $S=T=20$, $a=b=2$ and $N=800$.
The agreement is good, allowing for the errors inherent in the Monte Carlo results.
For interest, the final column shows the Poissonian probability obtained if the cells were treated as independent.
In this particular case, the result of assuming independence of the cells would be a fairly consistent overestimate of
the significance of the 'source' by about '$3\sigma$'.
The precise amount of underestimate of the chance probability will depend on the number of elemental cells in the
scanning window.
Finally, the treatment of \cite{kn:alm} has been extended to other shapes of scanning window, such as circular.

\subsubsection{Summary}
In summary, the 2-D Scan Statistic is a preferred general statistic for those cases where events are located 
randomly on a plane, within fixed bounds, and where there is no \emph{a priori} expectation such as a known source with 
known instrumental spread function.
In most practical situations a good approximation is obtained by using equation~\ref{eq:alm2d}.

%           MAIN SECTION                   BAYESIAN METHODS                      MAIN SECTION 
%           MAIN SECTION                   BAYESIAN METHODS                      MAIN SECTION 
%           MAIN SECTION                   BAYESIAN METHODS                      MAIN SECTION
 
\section{Bayesian Methods}
\label{sec:bayes}

\subsection{Introduction}

For many workers in cosmic rays, Bayesian methods are relatively novel and the following section attempts to 
summarise the main ideas and methods.
A much fuller development of the ideas discussed below is given by Loredo\cite{kn:loredo90,kn:loredo92,kn:loredo94}.

\subsubsection{Statistics}
The term 'statistics' arises from the concept of a \emph{statistic}. 
A statistic is a number derived from observed data and which obeys certain rules, some of which depend on a hypothesis
about the system under observation, some of which are extraneous. 
From this number, one can say how likely it is that the data was drawn from a population obeying rules specified by the 
particular hypothesis, assuming that all extraneous quantities are allowed for.
From that, by an inversion of logic, it is inferred how likely is the hypothesis.
In many cases, a particular statistic is used because the experimental results appear to be presented, or may be rearranged to
be presented, in a form which allows an easy calculation of that statistic.
An example is the epoch-folding of time-tagged photon times above, followed by \chisq\ calculated from the binned
phases. 
As was pointed out in that example, some arbitrary choices had to be made which rendered the results unsatisfactory.
Also, the aspects of the experimental data which were used to calculate a statistic may not be all that are available,
or the most discriminating aspects.
This should, with careful design, be evident from a consideration of the statistic's 'power' but not necessarily.
It is the claim of Bayesians that such problems are inherent in 'classical' statistics and derive from a misunderstanding
of the meaning of Probability.

\subsubsection{The Meaning of Probability}

There were at the beginnings of the subject, and still are, two schools of thought.
The first school maintains that the term \emph{probability} is a statement of the frequency of occurrence of data, 
such as that taken in a very large number of repeats of an experiment, under the assumption that random factors 
are at work causing the possibility that the results could be different every time. 
Take, as an example, coin-tossing: the probability of heads is obviously 0.5 in a single toss. 
Everyone would agree that, assuming no trickery, an unbiased coin would land equally likely as 'heads' or 'tails'. 
But there are forces at work which affect the way a coin would land - all amenable to analysis.
In fact a coin-tossing machine could be made which obtained 'heads' or 'tails' every time. 
We regard coin-tossing as a random activity only because we expect humans to apply unconscious variability to the force and 
direction of the flip which is much greater than that needed in the initial conditions to obtain one more extra turns before 
landing.
This illustrates an important point: unless the hypothesis is clear and specifies all pertinent factors there could be an 
apparent randomness.  
That is not to say that true randomness does not exist, only that it is often used as an alibi for lack of knowledge or 
precision in stating the experimental conditions.

An alternative definition of \emph{probability} is 'a measure of belief in a certain hypothesis'.
This is, and was, a much easier idea to grasp but one which was felt from an early date not to be capable of exact or 
scientific analysis. 
One consequence of this idea is that for a unique set of data, perhaps taken on a naturally-occurring phenomenon, 
the idea of a very large set of repeated experiments to plot out the 'frequency distribution necessary to use a 
'statistic' was unrealistic.
It is this definition which underlies Bayesian thought, and indeed is the definition which more closely accords with
the questions for which measurements are made. 
Interestingly, this meaning of probability explains the frequency version as a special case using de Finetti's representation 
theorem for exchangeable sequences of events\cite{kn:goldstein}.

The main difference between the two philosophical approaches is how the data are related to the hypotheses.
\begin{enumerate}
\item The 'Frequentist' approach:
We obtain $P(D\mid H)$, the conditional probability of obtaining the observed data, given a particular hypothesis. 
The hypothesis is frequently a model $M$ which has a parameter space $\theta$ and $P(D\mid M,\theta)$ is the 
'sampling distribution' for the data, given the model.
A frequently met hypothesis is the 'null' hypothesis $H_{o}$ in which the parameters are set to zero.
For any hypothesis, a statistic is formed which is 'locally most powerful' or even better 'uniformly most powerful' 
and the probability of observing the data is assigned from a knowledge of the statistic's distribution function.
This is now used to give a range of values in which the value of a statistic may fall by chance, with given probability
(i.e. frequency).
As an interesting aside, it is most usually the case for continuous measures, and frequently for discrete measures, for a
range of values of the statistic, including that observed (but also including many values \emph{not} observed), to be
used to derive the probability.
This is frequently performed by integrating $P(D\mid M,\theta)$ over the sample (data) space.
That is to say, the probability of a hypothesis is determined by the data taken, plus a whole range of values of
data which were \emph{not} observed.
This curious situation is not often questioned by ordinary users of 'statistics'.

\item The 'Bayesian' approach:
   We obtain $P(H\mid D)$, the probability of a hypothesis, given the data - apparently a more difficult matter.
   However Bayes Theorem gives: 
\begin{eqnarray*}
            P(H\mid D)P(D)  & = & P(D\mid H)P(H) \mbox{leading to}
\\ 
          P(H\mid D)     & = & P(D\mid H) P(H)/P(D)
\end{eqnarray*}
   $P(D\mid H)$ is the likelihood function, 
   $P(D)$ is the global likelihood, usually treated as an ignorable normalising constant,
   $P(H)$ is the 'prior' probability of the hypothesis. 
In addition to the extra terms not used in frequentist analysis, a crucial difference between the approaches is that
Bayesian methods would integrate the likelihood function over the parameters space, rather than the sampling (data) space. 
Bayesian methods have been criticised for the inclusion of an apparently subjective quantity $P(H)$ but a trivial
example demonstrates that frequentist analyses are not free from this. 
Frequentists would determine if the hypothesis $H_{o}$ of a histogram having all the cells identical were true by taking 
\chisq\ as a statistic. 
They would use no prior information or knowledge. 
But we know that histogram cells cannot contain negative numbers, and so some relevant background information is ignored 
when using \chisq. 
\end{enumerate}

\subsubsection{An Example}

 An example of the different approaches is an experiment in which a coin is tossed $N$ times. 
It lands heads $H$ times. The question is: is it biased?
In the Frequentist approach a hypothesis (the 'null' hypothesis) is formed that the coin is unbiased and that the result is 
a function of randomness only. 
A sufficiently low probability which is obtained for a suitable statistic would be evidence that the 'null hypothesis' 
should be abandoned.
The binomial distribution describes the result of such discrete, bounded experiments.
The probability of $H$ heads in $N$ tosses is $0.5^{H} \times 0.5^{N-H} \times C_{H}^{N}$. 
One then calculates the probability of obtaining $0,1,2,\ldots H-1$ heads and add them to the probability of $H$ heads and 
say: 'if the null hypothesis is true, getting $H$ \emph{or fewer} heads in $N$ tosses can occur due to chance, 
with a given probability. 
This could be interpreted as some evidence against the 'null hypothesis', hence evidence that the coin is biased.

\subsubsection{Stopping Rules}

Setting aside for the moment the fact that we did not see $0,1,2,\ldots H-1$ heads, the last conclusion supposes that the
coin was tossed, irrespective of the result, $N$ times and that the number of heads $H$ was the random variable. 
But suppose the coin was actually tossed by a person until $H$ heads were obtained, and that happened to occur after $N$ 
tosses. 
In this case the number of tosses $N$ is the random variable and the number of heads $H$ is fixed. 
The probability is then derived by combining the individual probabilities of obtaining $H$ heads from $h \geq H$ tosses,
and may be significantly different from the first probability calculated.

Loredo\cite{kn:loredo92} gives a more apposite example: a theorist predicts that $f=10\%$ of the stars in a cluster should
be of type \textbf{A}.
An observer reports 5 stars of type \textbf{A} out of 96 observed.
The theorist calculates as follows: $N=96$ and $f=0.1$ gives $9.6$ predicted type \textbf{A} stars.
The probability of the value of $\chi^{2}=2.45$ of this information is $P=0.073$, which is acceptable at the $95\%$ level.
The observer however decided in advance to stop when he had found 5 stars of type \textbf{A}.
The expected value of $N$ is then $5/0.1=50$ with variance $5(1-f)/f^{2}=450$.
The probability of the value of $\chi^{2}=4.7$ of this information is $P=0.032$, which is \emph{not} acceptable at the 
$95\%$ level.
This ambiguity arises because of the stopping rule used by the experimenter that is - what data sets \emph{might}
have been observed.

The stopping rule can therefore be important in classical statistical analysis, and ignorance of the actual rule used
may lead to an erroneous or at least ambiguous conclusion. 
Knowledge of the exact stopping rule is less important in Bayesian analysis, but is valuable in particular when it 
contains useful information about the unknown quantities.
In other cases, the stopping rule could be important if the existence of some data is unknown to the analyser, perhaps 
because its analysis did not show it to be significant and it was suppressed by the experimenter.
The message from this example is that for Frequentist analysis to be possible, an experiment must be precisely defined 
and if the execution is different in any way from the plan, the data could be worthless.

\subsubsection{Conclusion}
In summary, frequentist methods establish $P(D\mid \theta M)$ as the sampling distribution of the data, given a model
$M$ with parameters $\theta$ and perform integration over the data space.
Bayesian methods start with the same function $P(D\mid \theta M)$ but treat it is a likelihood with integrations
performed over the parameter space of the model.
In particular, parameters which are necessary for the specification of the model but are not of interest (for example
the phase when looking for a periodic signal) are integrated out, or marginalised.

In Bayesian theory, the notion of a 'random variable' is absent so ambiguity does not arise for many types of 
stopping rule and there is no need for a 'reference set' of hypothetical data.
This state of affairs results from the need in Bayesian methods to be specific about all the hypotheses, or to 
integrate away any unspecifiable variable.
Taking again the example of a histogram, and the question of whether its cells are consistent with uniformity using
\chisq\.
If the null hypothesis is the only hypothesis available, the use of \chisq\  is as a 'goodness-of-fit' test for the 
supposition of flatness. 
The number observed in the $i^{th}$ bin of $n$ bins is $x_{i}$. 
The number expected in each bin, under the null hypothesis, is $avg=\sum_{i=1}^{n}x_{i}/n$ and $\chi^{2}=
\sum_{i=1}^{n}(x_{i}-avg)^{2}/avg$, assuming $avg$ is large enough (usually 10 or so) for asymptotic normality.
The probability of \chisq\ for $n-1$ degrees of freedom is interpreted as supporting or otherwise the null hypothesis. 
This statistic suffers from a major problem in that it ignores information - the order of the bins may be significant, 
and so it implicitly assumes a class of alternative models in which the order is unimportant.
This can be partially rectified by applying an independent test which is only determined by the order of the bins - 
the Run Test. 
This is only applying a patch, since the Run test is most powerful against monotonicity and not other patterns. 
Frequentists acknowledge this problem in general by using the idea of the power of a statistic, that is its ability 
to identify correctly a true model from a particular alternative.
Both approaches have subjective factors: Bayesian in assigning prior probabilities to hypotheses, Frequentist in the 
notion of randomness and its applicability in a mathematical sense to cover for a lack of knowledge of the exact 
experimental conditions.
A consequence of this is that different experts in both fields may come to different conclusions given the same data.
Another way of putting this is that the result of analysing data will be a conclusion within a range, depending on
(a) the Bayesian priors, or (b) the estimate of the degrees of freedom and unknown factors.

%         Sub Section         BAYESIAN ON/OFF

\subsection{Bayesian On/Off Analysis}

The Bayesian ideas in the above section have recently been applied to the ON/OFF problem treated earlier.
As in all Bayesian analyses, some judgement must be made of the priors to be used, but in the cases discussed here 
the results do not depend critically on how these priors are chosen.

An initial Bayesian analysis of the problem of detecting a source in an ON/OFF counting experiment has been given by 
Loredo\cite{kn:loredo90}.
Using the same notation as in the ON/OFF section above, the probability of the background rate $b$ (the posterior density) 
from the OFF-source data is:
\[p\left(b\mid N_{OFF}\right)  =  p\left(N_{OFF}\mid b\right)\frac{p\left(b\right)}{p\left(N_{OFF}\right)}\]
The Poisson likelihood for $N_{OFF}$ is:
\[p(N_{OFF}\mid b)=\frac{(bT)^{N_{OFF}}e^{-bT}}{N_{OFF}!}\] 
The parameter $b$ is unknown and so the 'prior' probability would appear to be a matter of guesswork.
If the range of $b$ were pre-specified in some non-arbitrary way, at least the scale of $b$ would be known, and a flat
prior would be reasonable.
If even the scale of $b$ is unknown, the 'least informative' prior for $b$ is $p(b)=1/b$, which is uniform in $\log b$, 
and then
\[p(N_{OFF})=\frac{T^{N_{OFF}}}{N_{OFF}!}\int_{0}^{\infty}b^{N_{OFF}-1}e^{-bT} db\]
This leads to
\[p\left(b\mid N_{OFF}\right)  =  \frac{T_{OFF}\left(bT_{OFF}\right)^{N_{OFF}-1}e^{bT_{OFF}}}{\left(N_{OFF}-1\right)!}\]
Note that the expectation of the background $\hat{b}=N_{OFF}/T$ and that the assumption of $p(b)=1/b$ does not strongly 
affect the result, Loredo pointing out that a prior uniform in $b$ only marginally alters the expectation 
$\hat{b}=(N_{OFF}+1)/T$.
The joint probability of the background rate $b$ and a source rate $s$, given $N_{ON}$ and $N_{OFF}$,  is:
\[p\left(sb\mid N_{ON}\right)=p\left(s\mid b\right)p\left(b\right)\frac{p\left(N_{ON}\mid sb\right)}{p\left(N_{ON}\right)}\]
The probability of the source rate $s$ is obtained by marginalising $b$, that is $p\left(s\mid N_{ON}\right)=\int p
\left(sb\mid N_{ON}\right) db$:
\begin{equation}
p\left(s\mid N_{ON}\right)=\sum_{i=1}^{N_{ON}}C_{i}\frac{ T_{ON}\left(sT_{ON}\right)^{i-1}e^{-sT_{ON}} }{ \left(i-1\right)
! }
\label{eq:bayesonoff}
\end{equation}
where
\[C_{i}=\frac{ \left(1+\frac{1}{\alpha}\right)^{i}\frac{\left(N_{ON}+N_{OFF}-i-1\right)!}{\left(N_{ON}-i\right)!}}{ 
\sum_{j=1}^{N_{ON}}\left(1+\frac{1}{\alpha}\right)^{j}\frac{\left(N_{ON}+N_{OFF}-j-1\right)!}{\left(N_{ON}-j\right)!}}\]
This result is formally correct for all positive values of $N_{ON}$ and $N_{OFF}$ and is particularly useful for small
values when the asymptotic treatments fail.

Its main value is to illustrate the completely different approach and result of the application of Bayesian ideas.
However, there are some computational problems for values of $N_{ON}$ and $N_{OFF}$ which exceed $\sim 100$.
For values of $N_{ON}$ and $N_{OFF}$ which are less than $\sim 100$ evaluation of equation~\ref{eq:lambda} and equation
~\ref{eq:bayesonoff} shows small differences in the derived probabilities.

%         Sub Section         BAYESIAN BURSTS
%         Sub Section         BAYESIAN BURSTS
\subsection{Bayesian Change Point Analysis - Bursts}

A recent paper by Scargle\cite{kn:scargle98} has used Bayesian methods to analyse structure in photon counting data. 
It is worth noting that the ON/OFF problem dealt with above is a special case of change point analysis, where there is 
only one change point and its location is known in advance.
The principles are the same as those outlined above, with the added simplicity of having simpler alternatives to the uniform 
model. 
The uniform counting rate model $M_{1}$ assumes a constant intensity over a particular time interval $T$. 
An alternative model $M_{2}$ has the interval $T$ broken into two regions $T_{1}$ and $T_{2}$, $T=T_{1}+T_{2}$, each with 
a different counting rate. 
In general, a model $M_{k}$ may be constructed with $k$ regions. 
Bayes Theorem give the probability of a model
\[p\left(M_{k}\mid D,I\right)=\frac{p\left(D\mid M_{k},I\right)p\left(M_{k}\mid I\right)}{p\left(D\mid I\right)}\]
Dropping the explicit appearance of the background information $I$, the odds ratio $O_{kj}$ between two competing models 
$M_{k}$ and $M_{j}$ is then
\[\frac{p(M_{k}\mid D)}{p(M_{j}\mid D)}=\frac{p(D\mid M_{k})p(M_{k})}{p(D\mid M_{j})p(M_{j})}\]
The parameter $\theta$ or vector of parameters $\vec{\theta}$ of the model $M_{k}$ enter when $p(D\mid M_{k})$ is 
calculated
\[p(D\mid M_{k}) = \int p(D\mid \vec{\theta},M_{k})p(\vec{\theta}\mid M_{k})d\vec{\theta}\]
The odds ratio $O_{kj}$ is then
\begin{eqnarray}
 O_{kj} & = & \frac{p(M_{k}\mid D)}{p(M_{j}\mid D)} \nonumber
\\ & = & \frac{p(M_{k})}{p(M{j})}\frac{\int p(D\mid \vec{\theta},M_{k})p(\vec{\theta}\mid M_{k})d\vec{\theta} }
{\int p(D\mid \vec{\theta},M_{j})p(\vec{\theta}\mid M_{j})d\vec{\theta}} \nonumber
\\ & = & \frac{p(M_{k})}{p(M{j})} \frac{\mbox{$\mathcal{L}$}(M_{k},D)}{\mbox{$\mathcal{L}$}(M_{j},D)}
\end{eqnarray}                                                                    
where $\mbox{$\mathcal{L}$}(M_{k},D)$ is the global likelihood of model $M_{k}$.

For the constant-rate model $M_{1}$, $N$ events arrive in a time $T$ which is treated as being divided into $M$ intervals 
of duration $\delta t$, the justification being that photon counting apparatus always has a resolving time. 
Note that the number of events in any particular interval $\delta t$ can be $0$ or $1$ only.
The author shows that the global likelihood for this constant-rate model of Time Tagged Events (TTE) is
\[\mbox{$\mathcal{L}$}(M_{1}\mid TTE)=\frac{\Gamma(N+1)\Gamma(M-N+1)}{\Gamma(M+2)}\]
If the data is time-binned into $M$ equal bins, but such that any number of events may occur in any bin, given an overall 
rate $\lambda = N/T$ and mean number per bin of $\mu=\lambda T/M$, the global likelihood is:
\[\mbox{$\mathcal{L}$}(M_{1}\mid Binned)=\frac{\Gamma(N+1)}{(M+1)^{N+1}}\]
Note that the bins are fixed and may not be scanned to maximise $\mathcal{L}$.

The alternative model $M_{k}$ has a likelihood which is the product of the likelihoods of the individual constant-rate 
regions of $T$. 
For a two-rate model with the time of the change of rate being $t_{cp}$
\begin{eqnarray*}
\mbox{$\mathcal{L}$}(M_{2}\!\!\mid\!\!D) & = &\int\!\!dt_{cp}\int\!\!d\Lambda_{1}\int\!\!d\Lambda_{2} p_{cp}(t_{cp}) 
\times p[D_{1}\!\!\mid\!M_{1}(\Lambda_{1},\!T_{1})]p(\Lambda_{1}) 
\\
&   & \times p[D_{2}\!\!\mid\!M_{2}(\Lambda_{2},\!T_{2})]p(\Lambda_{2})
\end{eqnarray*}
where $\Lambda=\lambda \delta t$, $P(\Lambda)$ is the prior for the rate $\Lambda$ and $P_{cp}$ is the prior for the 
change-point time $t_{cp}$.
For time-tagged data with resolution $\delta t$ the integrals are sums and the change-point location is $m_{cp}\delta t$.
Since the change-point can be tested only at the arrival time of a photon, the photon number of the change-point $n_{cp}$ 
is used as an index. The number of events in the first section, up to the change-point, is $N_{1}=n_{cp}$, $N_{2}=N-N_{1}$ 
and $M_{1}=m_{n_{cp}}$
The global likelihood is then
\begin{eqnarray*}
\mbox{$\mathcal{L}$}(M_{2}\mid D) & = &\sum_{n_{cp}}\frac{\Gamma(n_{cp}\!\!+1)\Gamma(m_{n_{cp}}\!\!-n_{cp}\!+1)}{
\Gamma(n_{cp}\!\!+2)} \nonumber
\\
 &  & \times \frac{\Gamma(N\!-n_{cp}\!\!+\!1)\Gamma(m_{N-n_{cp}}\!-(N\!-n_{cp})+1)}{\Gamma(N\!-n_{cp}\!+2)}\Delta t_{n_{cp}}
\end{eqnarray*}

The paper\cite{kn:scargle98} gives a coding in a popular mathematical package to implement the above ideas.

%         Sub Section         BAYESIAN PERIODICITY
%         Sub Section         BAYESIAN PERIODICITY

\subsection{Bayesian Periodicity Analysis}

\subsubsection{Introduction}

Frequentist statistical theory allows more than one test to be applied to any situation.
Any statistic, or function of the data, may be defined and the 'best' is selected depending on its 'power' or likelihood 
of selecting the 'correct' hypothesis.  
One of the problems of the frequentist approach to looking for evidence of periodicity is that, in the absence of a 
specific light curve, the alternative hypothesis (to one of uniformity in the phase distribution) is unknown and the 
power of a statistical test is difficult to specify except for a narrow class of alternative light curves. 
The Rayleigh statistic, $Z_{1}^{2}$, is powerful only for the fundamental period and is formally the most powerful test 
for alternatives to uniformity from the Von Mises distribution - the circular equivalent of the Gaussian on the line.
The $Z_{n}^{2}$ test allows the addition of $n-1$ harmonics but needs a protocol to decide when to stop adding harmonics 
and therefore degrees of freedom (the $H$ test mentioned above suggests such a protocol).
Finally, Protheroe's test is powerful for very narrow light curves. 
Each could be tried in succession to look for evidence of periodicity, but a method which is indifferent to the shape 
of the light curve, without any penalty, would be of great advantage.

\subsubsection{Gregory \& Loredo Method}

Such a method based on Bayesian analysis, is claimed by Gregory and Loredo\cite{kn:gregory}. 
The essence of the method is to compare a uniform model for the distribution in phase at a trial frequency with a 
periodic model. 
The great difference between this and other methods is how the periodic model is proposed and how the necessary 
uncertainties and their associated 'degrees of freedom' of classical theory are accounted for. 
In particular, since an arbitrary postulated light curve may be of any shape, the method automatically applies Ockham's 
razor, in that models with fewer variables are automatically favoured unless the evidence from the data more than 
compensates. 
More complicated light curves (not necessarily with small number of harmonics, a $\delta$-function is uncomplicated in 
this context) are penalized for their greater complexity.

Bayes Theorem is used to compare the probabilities of two parameterised models of the phase distribution. 
In the notations of the authors, the probability that a model $M$ describes the data, given the data $D$ and any 
background information $I$ is 
\begin{equation}
p\left(M\mid D,I\right) = p\left(M\mid I\right) \frac{p\left(D\mid M,I\right)}
{p\left( D\mid I\right)}
\end{equation}
The first term on the right, $p\left(M\mid I\right)$, is the prior probability of the model $M$, which may seem to be 
subjective but may be estimated in some cases from the permissible range of the parameters. 
The numerator in the second term, $p\left(D\mid M,I\right)$, is the sampling probability of the data $D$, or the 
likelihood of the model $M$. The denominator, $p\left( D\mid I\right)$, is the global likelihood of the entire class of 
models. 
If the model contains a parameter $\theta$, or in the case two or more parameters a vector $\vec{\theta}$, the 
likelihood of the model can be calculated:
\[p\left (D\mid M\right)=\int_{\vec{\theta}}p\left( D\mid \vec{\theta},M\right)p\left(\vec{\theta} \mid M\right)\]

For time-tagged photon data with $N$ events detected over a time $T$, the probability of $D$ for a particular rate model 
$r(t)$ can be calculated. 
For the time $T$ divided into very small intervals of length $\Delta t$, the probability of $n$ events in $\Delta t$ is:
\[p_{n}=\frac{\left[r(t)\Delta t\right]^{n}e^{-r(t)\Delta t}}{n!}\]
If $\Delta t$ is small enough for $p_{i}=0, i\geq 2$ then the sequence of $T/\Delta t$ time samples will contain $N$ 
containing one event and $Q=T/\Delta t - N$ containing no event. 
The likelihood is then:
\[p\left(D\mid r,I\right)=\prod_{i=1}^{N}p_{1}(t_{i})\prod_{k=1}^{Q}p_{0}(t_{k})\]
Using $p_{0}(t)=e^{-r(t)\Delta t}$ and $p_{1}(t)=r(t)\Delta te^{-r(t)\Delta t}$ the likelihood function is
\[p\left(D\mid r,I\right) = \Delta t^{N}\left[\prod_{i=1}^{N}r(t_{i})\right]\exp\left[-\sum_{k=1}^{N+Q}r(t_{k})\Delta 
t\right]\]

In the case of a periodic model, the non-uniformity in phase is characterised by the varying contents of the phase bins. 
Although the number of phase bins needed to detect any light curve and the origin of phase are unknowns, these will be 
marginalised or integrated out. 
If there are $m$ phase bins the average rate $A=\frac{1}{m}\sum_{j=1}^{m}r_{j}$ and the fraction of the total rate per 
period in phase bin $j$ is $f_{j}=\frac{r_{j}}{mA}$.
The likelihood function is shown to reduce to
\[P\left(D\mid \omega,\phi,A,{\bf f},M_{m}\right)=\Delta t^{N}(mA)^{N}e^{-AT}\left(\prod_{j=1}^{m}f_{j}^{n_{j}}\right)\]
where $\omega$ is the postulated angular frequency, $\phi$ the starting phase, ${\bf f}$ the set of $m$ values of $f_{j}$ 
and $n_{j}$ being the number of events occurring in bin $j$.

The joint prior density for the parameters $\omega,\phi,A,{\bf f}$ is
\[p\left(\omega,\phi,A,{bf f}\mid M_{m}\right)=p\left(\omega\mid M_{m}\right)p\left(\phi\mid M_{m}\right)p\left(A\mid 
M_{m}\right)p\left({\bf f}\mid M_{m}\right)\]
The prior densities are:
\begin{enumerate}
\item $p(\phi \mid M_{m})=1/2\pi$,  this assumes that any starting phase is equally likely,
\item $p(A\mid M_{m})=1/A_{max}$, this assumes that $A$ does not change during the observation and any value of $A$ 
from $A=0$ to $A=A_{max}$ is possible,
\item $p(\omega \mid M_{m})=\/\omega \ln(\omega_{hi}/\omega_{lo})$, where $\left[\omega_{hi},\omega_{lo}\right]$ is 
a prior range for $\omega$,
\item $p({\bf f})=(m-1)!\delta\left(1-\sum_{j=1}^{m}f_{j}\right)$.
\end{enumerate}

The assignment of the priors of the models themselves is all that is needed before comparing the likelihoods of the models. 
The two models are equally likely \emph{a priori} and so the prior likelihood of the non-periodic model 
($M_{1}$), $p(M_{1})=1/2$ and that for the periodic model ($M_{m}$, $m=2,m_{max}$), $p(M_{m}\mid I)=1/2\nu$ 
where $\nu = m_{max}-1$.

The final result for the odds $O$ against a uniform model of phase and in favour of a periodic model with phase and 
period unknown (a common case) is:
\begin{equation}
O=\frac{1}{2\pi\nu\ln\left(\omega_{hi}/\omega_{lo}\right)}\frac{N!(m-1)!}{(N+m-1)!}\int_{\omega_{lo}}^{\omega_{hi}}
\frac{d\omega}{\omega}\int_{0}^{2\pi}d\phi\frac{m^{N}}{W_{m}\left(\omega,\phi\right)} \label{eq:odds}
\end{equation}
where $W_{m}\left(\omega,\phi\right)$ is the number of ways that the set of $n_{j}$ observed counts can be made by 
distributing $N$ counts in $m$ bins:
\[W_{m}\left(\omega,\phi\right)=\frac{N!}{\prod_{j=1}^{m}n_{j}!}\]
and $n_{j}$, the number of events placed in the $j^{th}$ phase bin depends on $\omega$, $\phi$ and $m$.

If the period is known, this reduces to:
\begin{equation}
O(\omega)=\frac{1}{2\pi\nu}\frac{N!(m-1)!}{(N+m-1)!}\int_{0}^{2\pi}d\phi \frac{m^{N}}{W_{m}\left(\omega,\phi\right)}
\label{eq:odds2}
\end{equation}

\begin{figure}[ht]
\psfig{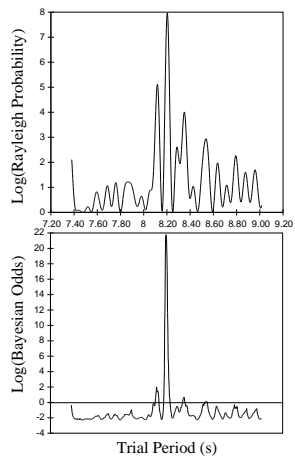}
\label{fig:bcomp}
\footnotesize
\caption{A comparison of Rayleigh Power and Bayesian Odds using random simulated data with a periodic signal at $P=8.2s$}
\end{figure}

In order to illustrate the difference between the information available from this treatment and from the Rayleigh
test, a data set has been generated containing time-tagged random events with a constant mean rate, plus a periodic
component.
The results are shown in figure~\ref{fig:bcomp}.
A particular point to note is that although the Rayleigh power is always positive, even for pure noise, in the case
of LOG(Bayesian odds), peaks do not become 'interesting' until they become positive. 
This is because the 'degrees of freedom' have been accounted for automatically and cause the offset seen in 
figure~\ref{fig:bcomp} so that peaks falling below $\log(Odds)=0$ are just those expected from noise.
It can be seen much more clearly in the Bayesian Odds diagram that there is only one significant peak, at the period
simulated. 

%%%%%%%%%%%%%%%%%%%%%%%%%%%%%%%%%%%%%%%%%%%%%%%%%%
\normalsize 
%%%%%%%%%%%%%%%%%%%%%%%%%%%%%%%%%%%%%%%%%%%%%%%%%%

The method has been used to detect a weak pulsar signal from SNR 0540-693 in ROSAT data, which could not be detected 
using a standard FFT technique\cite{kn:gregory96}. 
Moreover, the precision of determining the frequency was much higher for the Bayesian method than for $\chi^{2}$ 
using epoch-folding. 
The frequency precision of the latter is determined mainly by the duration of the data and is not strongly influenced 
by the number of photons.
Gregory and Loredo show\cite{kn:gregory96} that the Bayesian method obtains greater precision in parameter estimation 
with more photons.
The method has also been used to detect 1600 day modulation in the long-term radio emission of an X-ray binary, 
with very non-uniformly sampled data and a Gaussian noise of unknown magnitude \cite{kn:gregory99a,kn:gregory99b}.
There has been a recent independent use of a Bayesian method to calculate the upper limit to a pulsed flux at a
known period, independent of pulse width and pulse phase\cite{kn:VLAbayes}.

\section{Conclusions}

The hope of this review is that the more commonly met data analysis problems may be approached by the cosmic ray
worker with a more consistent and up to date approach.
There have been a number of advances in recent years in the tools, and more importantly in the methods, available to 
cosmic ray experimenters to ensure that the maximum use is made of hard-won data. 
The traditional statistical methods have resulted in a measure of agreement on the 'correct' way to look for sources
from ON/OFF data, change points (bursts) in 1- and 2-dimensions and in periodicity.
The application of these methods requires care to ensure that the 'degrees of freedom' are kept under control and   
properly accounted for: many of the criticisms of claimed sources have been based on the latter.

New Bayesian methods of testing hypotheses have recently been proposed.
A central theme of these methods is that classical methods often cloak ignorance in a way which distorts the results.
There are claimed to be significant benefits to the use of Bayesian methods which derive from the requirement to be 
absolutely specific about the hypotheses and the methodology of marginalising nuisance parameters.
In contrast to classical statistical methods, where various \emph{statistics} may be generated from the same data, each
with different assumptions, \emph{degrees of freedom} and {power}, Bayesian methods provide a framework for describing
completely the data and allow the direct comparison of specified hypotheses.
A practical result of the philosophical differences between the approaches is that, rather than relying on a relatively 
easy-to-use, pre-packaged test statistic, with the accompanying dangers of hidden degrees of freedom, a Bayesian
method requires the data interpreter to model the hypotheses precisely.
The obvious disadvantages of this are claimed to be more than compensated by the directness of the link between the
hypotheses and the data.
Bayesian methods may require some time to become accepted in the field, in that the methodologies and ideas have not
traditionally been part of the training of physicists; indeed may not have been as useful if physicists' training
in classical methods had been better.

\section{Acknowledgements}

The author would like to acknowledge useful discussions with P.S.Craig and M.Goldstein.

\end{document}